\documentclass[aps,prb,onecolumn,preprint,showpacs,superscriptaddress]{revtex4-1}
\usepackage{graphicx}
\usepackage{dcolumn}
\usepackage{bm}
\usepackage{amssymb}
\usepackage{color}
\usepackage{amsmath}

\begin{document}
\title{Theory of diffusive light scattering cancellation cloaking}
\author{M. Farhat}
\email{mohamed.farhat@kaust.edu.sa}
\affiliation{Division of Computer, Electrical, and Mathematical Sciences and Engineering, King Abdullah University of Science and Technology (KAUST), Thuwal 23955-6900, Saudi Arabia}
\author{P.-Y. Chen}
\affiliation{Department of Electrical and Computer Engineering, Wayne State University, Detroit, Michigan 48202, USA}
\author{S. Guenneau}
\affiliation{Aix-Marseille Universit\'e, CNRS, Centrale Marseille, Institut Fresnel, Campus universitaire de Saint-J\'er\^ome, 13013 Marseille, France}
\author{H. Bagci}
\affiliation{Division of Computer, Electrical, and Mathematical Sciences and Engineering, King Abdullah University of Science and Technology (KAUST), Thuwal 23955-6900, Saudi Arabia}
\author{K. N. Salama}
\affiliation{Division of Computer, Electrical, and Mathematical Sciences and Engineering, King Abdullah University of Science and Technology (KAUST), Thuwal 23955-6900, Saudi Arabia}
\author{A. Al\`u}
\affiliation{Department of Electrical and Computer Engineering, The University of Texas at Austin, Austin, TX, 78712, USA}

\date{\today}

\begin{abstract}
We report on a new concept of cloaking objects in diffusive light regime using the paradigm of the scattering cancellation and mantle cloaking techniques. We show numerically that an object can be made completely invisible to diffusive photon density waves, by tailoring the diffusivity constant of the spherical shell enclosing the object. This means that photons' flow outside the object and the cloak made of these spherical shells behaves as if the object were not present. Diffusive light invisibility may open new vistas in hiding hot spots in infrared thermography or tissue imaging.
\end{abstract} 


\maketitle

\section{Introduction}

Invisibility cloaks that were introduced ten years ago \cite{leonhardt2006,pendry2006,schurig2006,guenneau2011} are undoubtedly the most popular application of artificial materials \cite{sihvola2007}, since they offer an unprecedented control over light trajectories \cite{pendry2006}. In order to make objects invisible to electromagnetic radiation, numerous methods have been suggested in the past few years. The first attempt can be traced back to Kerker and his consideration of core-shell dielectric bodies. He noted forty years ago that in the static regime, some combinations of dielectric functions lead to cancellation of the scattering cross-section 
of coated ellipsoidal and spherical inclusions \cite{kerker1975,chew1976abnormally}. Al\`u and Engheta further investigated the scattering cancellation technique and have shown that the use of plasmonic materials can lead to invisibility via the polarizability engineering of the coating shell \cite{alu2005,milton2006cloaking,engheta2007}. This technique has the advantage of being robust \cite{engheta2007} and experimentally realizable \cite{rainwater2012}. It has also many applications in sensing \cite{alu2009} or imaging \cite{alu2010}. Variants of this technique have been also investigated, e.g. by using plasmonic nano particles for optical cloaking \cite{muhlig2011,muhlig2013} or mantle covers \cite{alu2009mantle,farhat20133d}. Different methods based on homogenization have been also shown to lead to broadband cloaking of electromagnetic waves \cite{cai2007optical,farhat2008}. 

In the same vein, the scattering cancellation technique has been successfully adapted to other kinds of waves. In fact, Chen \emph{et al.} have shown that spherical objects can be made invisible to acoustic pressure waves by coating them with ultra-thin tailored shells with properly designed acoustic impedance \cite{chen2011,farhat2012}. The same technique has shown its potential even for the scalar elasticity domain (thin-plates), where waves obey the fourth order scalar biharmonic equation \cite{farhat2014platonic}, or thermal waves \cite{farhat2015thermal}. 

In this work, we consider the propagation of diffuse photon density waves (DPDW). These describe the propagation of light waves in a highly scattering medium, i.e. a turbid medium such as human body tissues, cloud or milk \cite{cheong1990review}. In this case, photons experience multiple scattering phenomenon before being absorbed by the medium or escaping it. Even though, the individual photons follow random paths, macroscopically, they behave like a photon density wave, described by a Helmholtz-like differential equation and experiencing usual wave phenomena such as refraction \cite{o1992refraction} scattering \cite{boas1993scattering}, or even cloaking based on transformation optics \cite{schittny2014invisibility,schittny2015diffuse}. We will propose thus to analyze the scattering of DPDW from core-shell structures and derive the conditions of scattering cancellation and resonant scattering regimes. The latter scenario may have promising applications in medical tissue imaging \cite{huang1991optical} by rendering objects more visible or improving the capabilities of HAMR technology \cite{kryder2008heat}.

\section{Dispersion relation of diffuse photon density waves} 

\begin{figure}[t]
\centering\includegraphics[width=0.95\columnwidth]{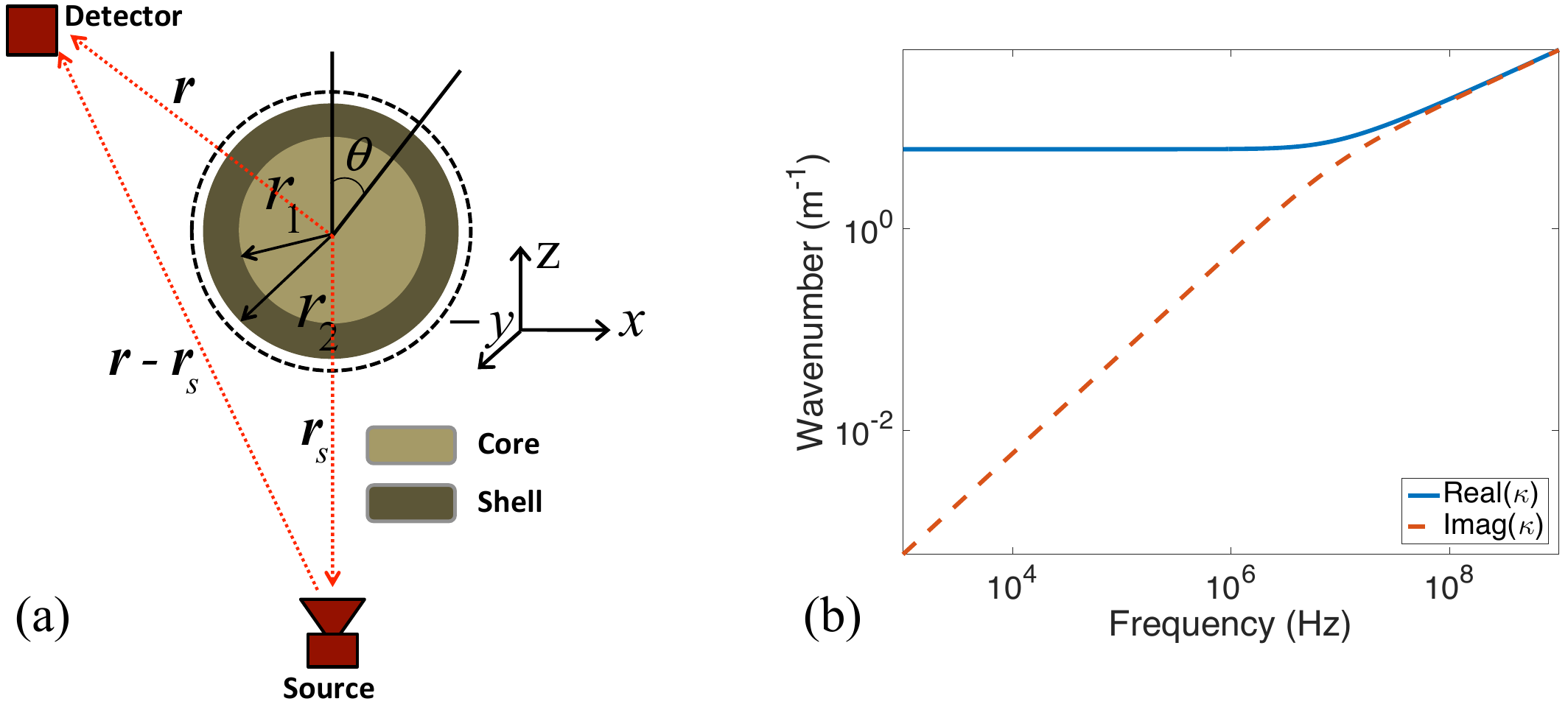}
\caption{Light diffusion scattering problem: Cross-sectional view of the light diffusion transfer scenario, with the object to cloak in the middle. (b) Dispersion relation of diffusion waves: Real and imaginary parts of the wavenumber $\kappa_0$ are given versus frequency.}
\label{fig1}
\end{figure}
Using the first principle of thermodynamics in a closed system, one can show that (in the absence of radiation and convection) the photon number density $\Phi({\bf r})$ or the photons fluence obeys the relation $\nabla\cdot {\bf j}+\partial\Phi/\partial t+\upsilon\mu_a\Phi+\upsilon\varsigma=0$. Here, ${\bf j}$, $\upsilon$, $\mu_a$, $\varsigma$, and $\Phi$ represent the photon current density (photons flow per unit surface per unit time), the speed of light in the diffusive medium, the absorption coefficient, the source of photons, and the photons fluence field, respectively. Using Fick's first law, i.e. the linear and instantaneous relation ${\bf j}=-D\nabla\phi$, one can derive that
\begin{equation}
\frac{\partial\Phi}{\partial t}+\upsilon\mu_a\Phi+\upsilon\varsigma=\nabla\cdot\left(D\nabla\Phi\right).
\label{diff_eq}
\end{equation}
The diffusivity is $D=\upsilon/[3(\mu_a+\mu_s')]\approx\upsilon/(3\mu_s')$, with $\mu_s'$ the scattering coefficient. The approximation $\mu_a\ll\mu_s'$ is called the $P_1$ approximation \cite{furutsu1994diffusion,frank2007approximate,heizler2010asymptotic} and is necessary to have the simplified form of Eq.~(\ref{diff_eq}). Assuming a piecewise constant diffusivity, the equation for $\Phi$ becomes in the time harmonic regime ($e^{-i\omega t}$)
\begin{equation}
\Delta\Phi+\left(\frac{i\omega-\upsilon\mu_a}{D}\right)\Phi=\upsilon\varsigma,
\label{diff_eq_harm}
\end{equation}
showing that these DPDW \emph{pseudo}-waves obey a Helmholtz-like equation $\Delta\phi+\kappa^2\phi=0$, with a complex \emph{pseudo}-wavenumber $\kappa$ related to the frequency through the dispersion relation $\kappa^2=(i\omega-\upsilon\mu_a)/D$. Using the parameters of water, i.e. a diffusivity $D_0=1.75\times10^6\, \textrm{m}^2/\textrm{s}$ and a lifetime of the photons $1/(\upsilon_0\mu_{a,0})=30\times10^{-9}\textrm{s}$, with the subscript 0 denoting the surrounding medium properties, the dispersion relation of photons is shown in Fig.~\ref{fig1}(b). In particular, one can notice that the real part $\Re(\kappa_0)$ tends asymptotical to the limit $\sqrt{\upsilon_0\mu_{a,0}/D_0}$ when the frequency goes to zero, and is much higher than the imaginary part, due to absorption. At higher frequencies ($\omega>>\upsilon_0\mu_{a,0}$), the real and imaginary parts are equal [since $\kappa_0^2$ is purely imaginary, as can be seen from the low frequency regime in Fig.~\ref{fig1}(b)].   
 
\section{Scattering cancellation modeling for DPDW}

In this section, we assume that generation of macroscopic waves (DPDW) from sinusoidally modulated point sources in a turbid medium is possible \cite{o1992refraction,tromberg1993properties}. These are spherical waves that propagate outwards from the point source, and are incident on a core-shell (spherical) structure, as schematized in Fig.~\ref{fig1}(a). The scattering (and absorbing) object (properties are indexed with subscript 1) is coated with a cloaking shell of tailored parameters (properties are indexed with subscript 2). It can be shown (as generally done for other kinds of waves \cite{alu2005}) that the fluence (or photons density) $\Phi$ can be found by constructing a general solution in the different domains [0, 1 and 2 as shown in Fig.~\ref{fig1}(a)] and applying the convenient boundary conditions. Without loss of generality, we assume the origin of the spherical coordinates to coincide with the center of the core-shell. 

\subsection{Heterogeneous scatterers}

Let us consider to begin, heterogeneous scattering objects, i.e. with finite values for $D_1$ and $\upsilon_1\mu_{a,1}$. Outside the structure (region 0), the general solution is the well-known superposition of incident and scattered waves, i.e. $\Phi^{0}=\Phi^{\textrm{inc}}+\Phi^{\textrm{scat}}$, where
\begin{equation}
\begin{aligned}
\Phi^{\textrm{inc}}(r,\theta,\varphi) ={} & \frac{\upsilon\varsigma}{4\pi D_0|{\bf r}-{\bf r}_s|} e^{i\kappa_0|{\bf r}-{\bf r}_s|} \\
     ={} & i \frac{\upsilon\varsigma\kappa_0}{D_0}\sum_{l=0}^{\infty}j_l(\kappa_0r_{<})h_l^{(1)}(\kappa_0r_{>})\sum_{m=-l}^{l}Y_{l,m}^{*}(\theta_s,\varphi_s)Y_{l,m}(\theta,\varphi),
\end{aligned}
\label{phi_inc}
\end{equation}
is the spherical wave created by the source and incident into the structure of Fig.~\ref{fig1}. Here ${\bf r}$ and ${\bf r}_s$ denote the positions of the detector and the source, respectively, and $r_{<}$ and $r_{>}$ is the smaller and larger of ${\bf r}$ and ${\bf r}_s$, respectively. The $^*$ denotes the complex conjugate. The wave scattered from the structure can be expressed as 
\begin{equation}
\Phi^\mathrm{scat}(r,\theta,\varphi)=\sum_{l,m}s_{l,m}h_l^{(1)}(\kappa_0r)Y_{l,m}(\theta,\varphi), \, r>r_2.
\label{phi_scatt}
\end{equation}
Inside the core (i.e. $r\leq r_1$), the solution is
\begin{equation}
\Phi^\mathrm{1}(r,\theta,\varphi)=\sum_{l,m}a_{l,m}j_l(\kappa_1r)Y_{l,m}(\theta,\varphi).
\label{phi_core}
\end{equation}
Inside the cloaking shell (i.e. $r_1<r\leq r_2$), the solution is
\begin{equation}
\Phi^\mathrm{2}(r,\theta,\varphi)=\sum_{l,m}[b_{l,m}j_l(\kappa_2r)+c_{l,m}y_l(\kappa_2r)]Y_{l,m}(\theta,\varphi).
\label{phi_shell}
\end{equation}
Here, $j_l$, $y_l$, and $h_l^{(1)}$ are spherical Bessel functions of the first, second kind and the Hankel function of the first kind, respectively and $Y_{l,m}$ are spherical harmonics. $\kappa_0$, $\kappa_1$ and $\kappa_2$ denote the complex wave numbers in the different regions of space (i.e. regions 0, 1, and 2, respectively). The coefficients of the developments ($a_{l,m}$, $b_{l,m}$, $c_{l,m}$, and $s_{l,m}$) can be determined using the following assumptions: (i) $\Phi$ is finite for ${\bf r}\neq{\bf r}_s$. (ii) When $r\rightarrow\infty$, $\Phi^0$ tends asymptotically to a spherically outgoing wave. (iii) The normal component of the flux is continuous, that is $D_1\partial\Phi^1/\partial r=D_2\partial\Phi^2/\partial r$ and $D_2\partial\Phi^2/\partial r=D_0\partial\Phi^0/\partial r$ for $r=r_1$ and $r=r_2$, respectively. (iv) $\Phi$ is continuous, i.e. $\Phi^1(r_1)=\Phi^2(r_1)$ and $\Phi^2(r_2)=\Phi^0(r_2)$.

Without loss of generality, and to simplify the calculations, we assume that the source is located on the $z$-axis (i.e. $\theta_s=\pi$ and $\phi_s=0$). This means that the terms for which $m\neq0$ are identically zero. These boundary conditions yield, therefore, the different coefficients. In particular, the scattering coefficients $s_l$ can be re-expressed as $s_l=-\psi_l/\left(\psi_l+i\chi_l\right)$. Here $\psi_l$ and $\chi_l$ are given by the complex-valued determinants
\begin{equation}
\psi_l=
\begin{vmatrix}
-j_l(\kappa_1r_1)& y_l(\kappa_2r_1)&j_l(\kappa_2r_1)&0\\
-D_1\kappa_1j_l^\prime(\kappa_1r_1)& D_2\kappa_2y_l^\prime(\kappa_2r_1)&D_2\kappa_2j_l^\prime(\kappa_2r_1)&0\\
0& y_l(\kappa_2r_2)& j_l(\kappa_2r_2)& \alpha_lj_l(\kappa_0r_2)\\
0& D_2\kappa_2y_l^\prime(\kappa_2r_2)& D_2\kappa_2j_l^\prime(\kappa_2r_2)& D_0\kappa_0\alpha_lj_l^\prime(\kappa_0r_2)
\end{vmatrix},
\label{psi_l}
\end{equation}
and
\begin{equation}
\chi_l=
\begin{vmatrix}
-j_l(\kappa_1r_1)& y_l(\kappa_2r_1)&j_l(\kappa_2r_1)&0\\
-D_1\kappa_1j_l^\prime(\kappa_1r_1)& D_2\kappa_2y_l^\prime(\kappa_2r_1)&D_2\kappa_2j_l^\prime(\kappa_2r_1)&0\\
0& y_l(\kappa_2r_2)& j_l(\kappa_2r_2)& \alpha_ly_l(\kappa_0r_2)\\
0& D_2\kappa_2y_l^\prime(\kappa_2r_2)& D_2\kappa_2j_l^\prime(\kappa_2r_2)& D_0\kappa_0\alpha_ly_l\prime(\kappa_0r_2)
\end{vmatrix},
\label{chi_l}
\end{equation}
with $\alpha_l=i(\upsilon\varsigma\kappa_0/D_0)j_l(\kappa_0|{\bf r}_s|)Y_{l,m}^{*}(\pi,0)$.

The scattering cross-section (SCS) $\sigma^\mathrm{scat}$ is a measure of the overall visibility of the object to external observers. It is obtained by integrating the scattering amplitude and can be expressed as \cite{farhat2012}
\begin{equation}
\sigma^\mathrm{scat}=\frac{4\pi}{|\kappa_0|^2}\sum_{l=0}^{\infty}(2l+1)\frac{|\psi_l|^2}{|\psi_l+i\chi_l|^2}.
\label{scs_scatt}
\end{equation}
In the limit of small scatterers (long diffusion length $\kappa_0r_{1,2}\ll1$), only few scattering orders contribute to the SCS. Here, we assume that the first two orders ($l=0$ for the monopole, and $l=1$ for the dipole mode, unlike in the electromagnetic case, where the first dominant mode is the dipole one) are significant. In this scenario, one has $\sigma^\mathrm{scat}\approx 4\pi/(|\kappa_0|^2)\left(|s_0|^2+3|s_1|^2\right)$. Consequently, canceling these two modes, i.e. $\psi_0=0$ and $\psi_1=0$, will ensure that $\sigma^\mathrm{scat}\approx0$, and the \emph{diffusive} scattering from the object will be suppressed. Namely, the SCT conditions on the parameters of the cloaking shell $D_2$, $\upsilon_2\mu_{a,2}$, and $r_2$ are
\begin{equation}
\frac{\left(1+i\upsilon_2\mu_{a,2}/\omega\right)-\left(1+i\upsilon_0\mu_{a,0}/\omega\right)}{\left(1+i\upsilon_2\mu_{a,2}/\omega\right)-\left(1+i\upsilon_1\mu_{a,1}/\omega\right)}=\frac{\upsilon_2\mu_{a,2}-\upsilon_0\mu_{a,0}}{\upsilon_2\mu_{a,2}-\upsilon_1\mu_{a,1}}=\gamma^3,\, \textrm{for $\psi_0=0$},
\label{l=0}
\end{equation}
and
\begin{equation}
\frac{(D_0-D_2)(D_1+2D_2)}{(D_1-D_2)(D_0+2D_2)}=\gamma^3,\, \textrm{for $\psi_1=0$},
\label{l=1}
\end{equation}
with $\gamma=r_1/r_2$. The monopole SCT condition in Eq.~(\ref{l=0}), depends on the product of the photons' velocity and the absorption coefficient of the shell, and the ratio of radii of the object and the shell $\gamma$. Likewise, the condition in Eq.~(\ref{l=1}) depends only on the diffusivity coefficient of the shell and $\gamma$. By enforcing these two conditions, the total scattering from the spherical object can be suppressed in the limit of small scatterers. 

\subsection{Perfectly absorbing scatterers}

We consider here the case of perfectly absorbing spherical objects. The developments given in Eqs.~(\ref{phi_inc}),(\ref{phi_scatt}) and (\ref{phi_shell}) remain unchanged but the fluence field becomes identically zero inside the core, i.e. $\Phi^{1}=0$. Additionally, the boundary condition at $r=r_2$ is the same as in the previous sub-section, but at $r=r_1$, we must apply a zero partial flux boundary condition, i.e.
\begin{equation}
\frac{D_2}{2\upsilon_2}\frac{\partial\Phi^2(r,\theta,\varphi)}{\partial r}-\frac{1}{4}\Phi^2(r,\theta,\varphi)=0.
\label{partial_flux}
\end{equation} 
These new boundary conditions lead to the following scattering coefficients,
\begin{equation}
\psi_l=
\begin{vmatrix}
\frac{D_2\kappa_2}{2\upsilon_2}y_l^\prime(\kappa_2r_1)-\frac{1}{4}y_l(\kappa_2 r_1)&\frac{D_2\kappa_2}{2\upsilon_2}j_l^\prime(\kappa_2r_1)-\frac{1}{4}j_l(\kappa_2 r_1)&0\\
y_l(\kappa_2r_2)& j_l(\kappa_2r_2)& \alpha_lj_l(\kappa_0r_2)\\
D_2\kappa_2y_l^\prime(\kappa_2r_2)& D_2\kappa_2j_l^\prime(\kappa_2r_2)& D_0\kappa_0\alpha_lj_l^\prime(\kappa_0r_2)
\end{vmatrix},
\label{psi_l_abs}
\end{equation}
and
\begin{equation}
\chi_l=
\begin{vmatrix}
\frac{D_2\kappa_2}{2\upsilon_2}y_l^\prime(\kappa_2r_1)-\frac{1}{4}y_l(\kappa_2 r_1)&\frac{D_2\kappa_2}{2\upsilon_2}j_l^\prime(\kappa_2r_1)-\frac{1}{4}j_l(\kappa_2 r_1)&0\\
y_l(\kappa_2r_2)& j_l(\kappa_2r_2)& \alpha_ly_l(\kappa_0r_2)\\
D_2\kappa_2y_l^\prime(\kappa_2r_2)& D_2\kappa_2j_l^\prime(\kappa_2r_2)& D_0\kappa_0\alpha_ly_l^\prime(\kappa_0r_2)
\end{vmatrix}.
\label{chi_l_abs}
\end{equation}
In the limit of small scatterers, i.e. $\kappa_ir_j\ll1$, with $i,j$ referring to the different regions of space, we can approximate the scattering coefficient using the leading terms in the Bessel functions. For $l=0$, we have (to the leading order) 
\begin{equation}
\psi_0\approx\frac{D_2\kappa_2}{4\kappa_2^2r_2^2},
\label{l=0_abs}
\end{equation} 
meaning that in order to cancel the first scattering coefficient, we need to set $D_2=0$. For the next scattering coefficient, that is $l=1$, after some straightforward algebra, we can show that the condition that ensures that $\psi_1=0$ (to the leading order) can be given  
\begin{equation}
\frac{D_2-D_0}{2D_2+D_0}=\frac{\gamma^3}{2}.
\label{l=1_abs}
\end{equation} 

\subsection{Ultra-thin invisibility shells: mantle cloaking}

It has been recently shown that a patterned metasurface can produce significant cloaking efficiency using simpler and thinner geometries \cite{alu2009mantle,chen2011,farhat2012}. The setup of the scattering problem is identical to what was discussed earlier, except that the scattering reduction is achieved here by a 3D surface, instead of a thick shell. Our aim in the following is to show the possibility of drastically reducing the scattering from spherical heterogeneities by properly tailoring the surface impedance. There are two boundary conditions that should be satisfied at the surface of both spherical obstacle (on $r=r_1$) and the cloak ($r=r_2$). Across the boundary $r=r_1$, we have the same conditions as in sub-section 3.1 (i.e. continuity of $\Phi$ and the normal component of its flux $D_i\partial\Phi^i/\partial r$. On the boundary of the mantle cloak, however, we have to consider the diffusive surface impedance which implies a jump of the normal component of the flux. And the boundary condition becomes
\begin{equation}
\displaystyle
\left[\frac{1}{\mu_{s}'}\frac{\partial\Phi}{\partial n}\right]_{r=r_2^+}^{r=r_2^-}=\frac{3Z^{-1}_d}{\upsilon}\Phi|_{r=r_2},
\label{impedance}
\end{equation}
where we have used the continuity of the fluence field $\Phi|_{r=r_2^-}=\Phi|_{r=r_2^+}=\Phi|_{r=r_2}$, and where $Z_d=R_d+iX_d$ is the averaged surface impedance that relates the fluence to the its flux on the surface. This impedance is function of the geometry of the structure and the wavelengths of the excitation signals and can usually vary in a large range of values.
\begin{figure}[t]
\centering\includegraphics[width=0.9\columnwidth]{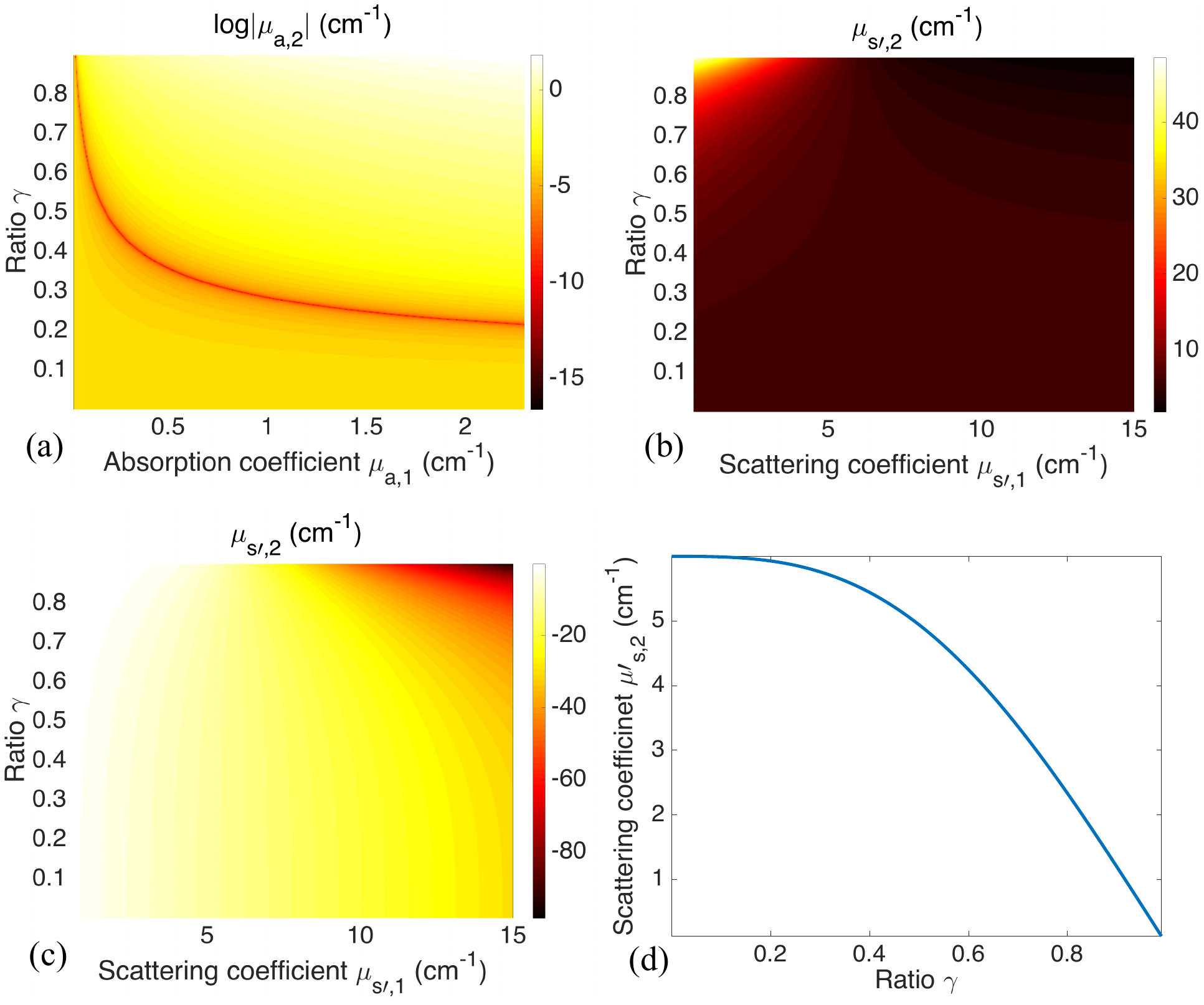}
\caption{(a) Contours of optimal absorption coefficient $\mu_{a,2}$ in logarithmic scale, versus the absorption coefficient of the object $\mu_{a,1}$ and the ratio $\gamma=r_1/r_2$. (b) Positive and (c) negative solutions of Eq.~(\ref{l=1}) giving contours of the scattering coefficient $\mu_{s,1}'$ and the ratio $\gamma$. (d) Scattering coefficient $\mu_{s,2}'$ given by Eq.~(\ref{l=1_abs}) for the case of perfectly absorbing obstacles.}
\label{fig2}
\end{figure}   

The $l$th spherical scattering order can be suppressed provided that the following determinant is canceled
\begin{equation}
\psi_l=
\begin{vmatrix}
-j_l(\kappa_1r_1)& y_l(\kappa_0r_1)&j_l(\kappa_0r_1)&0\\
-D_1\kappa_1j_l^\prime(\kappa_1r_1)& D_2\kappa_0y_l^\prime(\kappa_0r_1)&D_2\kappa_0j_l^\prime(\kappa_0r_1)&0\\
0& y_l(\kappa_0r_2)& j_l(\kappa_0r_2)& \alpha_lj_l(\kappa_0r_2)\\
0& y_l^\prime(\kappa_0r_2)+\eta y_l(\kappa_0r_2)& j_l^\prime(\kappa_0r_2)+\eta j_l(\kappa_0r_2)& j_l^\prime(\kappa_0r_2)
\end{vmatrix},
\label{psi_l}
\end{equation}
where $\eta=i\omega/(Z_dD_0\kappa_0\alpha_l)$ is a dimensionless function that accounts for the diffusive surface impedance. In the limit of large wavelengths, the spherical Bessel functions can be simplified and the approximate cloaking condition (for the reactance) can be written as
\begin{equation}
X_d=\frac{2\upsilon}{9\gamma^3\omega r_1}\left(\frac{\gamma^3}{\mu_{s,0}'}+\frac{\mu_{s,1}'+2\mu_{s,0}'}{\mu_{s,0}'(\mu_{s,0}'-\mu_{s,1}')}\right).
\label{impedance_lim}
\end{equation}
This clearly demonstrates that by properly choosing the diffusive surface reactance ($X_d$), it is possible, in the quasistatic limit, to suppress the few dominant scattering multipoles.

\section{Numerical modeling} 

\subsection{Ideal cloaking parameters}

The monopole SCT condition of Eq.~(\ref{l=0}), depends merely on the product of the speed of light and the absorption coefficient of the cloaking shell, and the ratio of radii $\gamma$. Similarly, the condition of Eq.~(\ref{l=1}) depends only on the scattering coefficient of the cloaking shell and $\gamma$. By enforcing these two conditions, one is able to cancel the total scattering from the spherical object, in the limit of small scatterers (dilute limit). It can be seen from Figs.~\ref{fig2}(a)-\ref{fig2}(c), the numerical solutions to Eqs.~(\ref{l=0}), (\ref{l=1}), where the variation of the absorption coefficient $\mu_{a,2}$ and the scattering coefficient $\mu_{s,2}'$ were plotted as function of the aspect ratio of obstacle to shell radius $\gamma$ and $\mu_{a,1}$ and $\mu_{s,2}'$, respectively. For the first case, given in Fig.~\ref{fig2}(a), solution of Eq.~(\ref{l=0}), one can see that the absorption coefficient of the shell ($\mu_{a,2}$), given here in logarithmic scale, takes positive and negative values, depending on $\gamma$ and the absorption of the object. The red thick line represents the curve obeying the equation $\gamma^3\mu_{a,1}/\mu_{a,0}=1$, meaning that $\mu_{a,2}=0$. The absorption coefficient takes thus negative (respectively positive near zero) values above (respectively below) this curve. For the second case, given in Figs.~\ref{fig2}(b) and~\ref{fig2}(c), solutions of Eq.~(\ref{l=1}), it can be seen that the required scattering coefficient of the shell $\mu_{s,2}'$ needs to be either positive [Fig.~\ref{fig2}(b)] or negative [Fig.~\ref{fig2}(c)], for varying $\gamma$ and $\mu_{s,1}'$, in order to satisfy Eq.~(\ref{l=1}) that has two solutions. However, for a small perfectly absorbing object, the scattering coefficient of the shell depends only on the ratio $\gamma$ and is always positive, varying from $\mu_{s,0}'$ to values close to zero, following Eq.~(\ref{l=1_abs}). 

\subsection{Canceling the SCS}

\begin{figure}[b]
\centering\includegraphics[width=0.95\columnwidth]{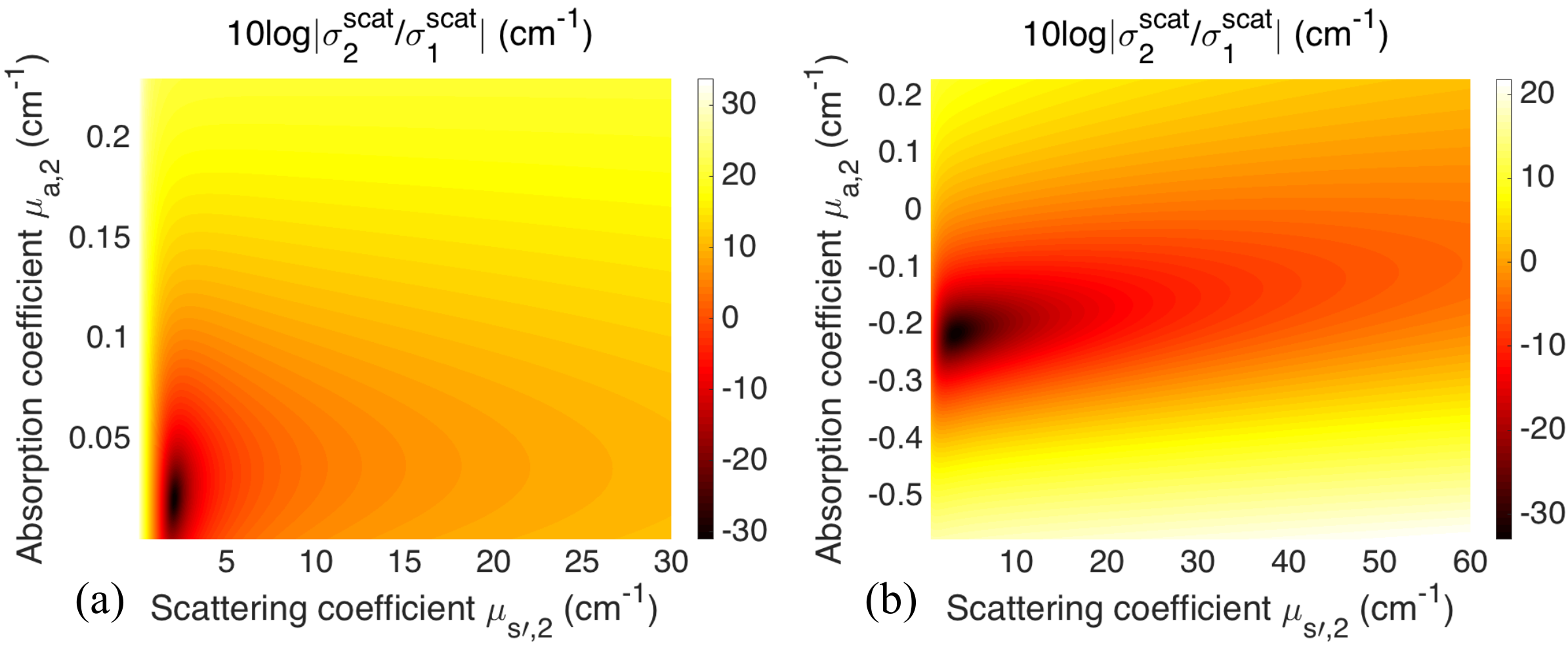}
\caption{Contours of the normalized SCS $\sigma^\textrm{scat}$ (to the SCS of the bare object) in logarithmic scale, versus the scattering coefficient $\mu_{s,2}'$ and the absorption coefficient $\mu_{a,2}$ for (a) a "scattering" object with $\mu_{s,1}'=2.5\mu_{s,0}'=15\textrm{cm}^{-1}$ and $\mu_{a,1}=\mu_{a,0}=0.023\textrm{cm}^{-1}$ and (b) an "absorbing" object with $\mu_{s,1}'=\mu_{s,0}'=6\textrm{cm}^{-1}$ and $\mu_{a,1}=6.5\mu_{a,0}=0.15\textrm{cm}^{-1}$. The white dots represent the positions of optimized scattering reduction, with a value of -35 dB.}
\label{fig3}
\end{figure}

Let us consider in this sub-section two particular scenarios, where scattering of DPDW from spherical heterogeneities is analyzed. The background, i.e. turbid medium has a scattering coefficient $\mu_{s,0}'=6\textrm{cm}^{-1}$ and an absorption coefficient $\mu_{a,0}=0.023\textrm{cm}^{-1}$ (the $P_1$ approximation is largely valid here). Two kinds of heterogeneities are considered here: a scattering object and an absorbing object. For the first kind (scattering), we have $\mu_{s,1}'=15\textrm{cm}^{-1}$ and $\mu_{a,1}=\mu_{a,0}$. For the latter (absorbing), we have $\mu_{s,1}'=\mu_{s,0}'$ and $\mu_{a,1}=0.15\textrm{cm}^{-1}$. The radii of both objects are $r_1=1.2\textrm{cm}$. The design turbid medium wavenumber is chosen as $\kappa_0r_1=0.5$. These two spheres are then coated with an invisibility shell of radius $r_2=1.125r_1=1.35\textrm{cm}$, and designed scattering and absorption coefficients $\mu_{s,2}'$ and $\mu_{a,2}$, respectively. 
We then compute the scattering cross-section of the total structure, using Eq.~(\ref{scs_scatt}), and normalize it with respect to the SCS of the respective bare objects. The resulting $\sigma^{\textrm{scat}}$ are plotted (in logarithmic scale) in Figs.~\ref{fig3}(a) and~\ref{fig3}(b) for the scattering and absorbing objects, respectively. By varying $\mu_{s,2}'$ and $\mu_{a,2}$ of the shell, we can observe that there are regions (dark) corresponding to huge scattering reduction, whereas light (white) regions correspond to increased scattering from the structure. For the scattering object [Fig.~\ref{fig3}(a)], values of $\mu_{s,2}'$ between 1.5 and 3.75 and $\mu_{a,2}$ between 0.01 and 0.045 are ideal for DPDW scattering cancellation (black spot). In particular, the minimum of $\sigma^\textrm{scat}$ is obtained for the coordinates (2.1, 0.028) that fits very well with the predictions of Eqs.~(\ref{l=0}), and~(\ref{l=1}). For the absorbing object [Fig.~\ref{fig3}(b)], values of $\mu_{s,2}'$ between 1 and 15 and $\mu_{a,2}$ between -0.15 and -0.28 are ideal for DPDW scattering cancellation (black spot). 
\begin{figure}[b]
\centering\includegraphics[width=0.9\columnwidth]{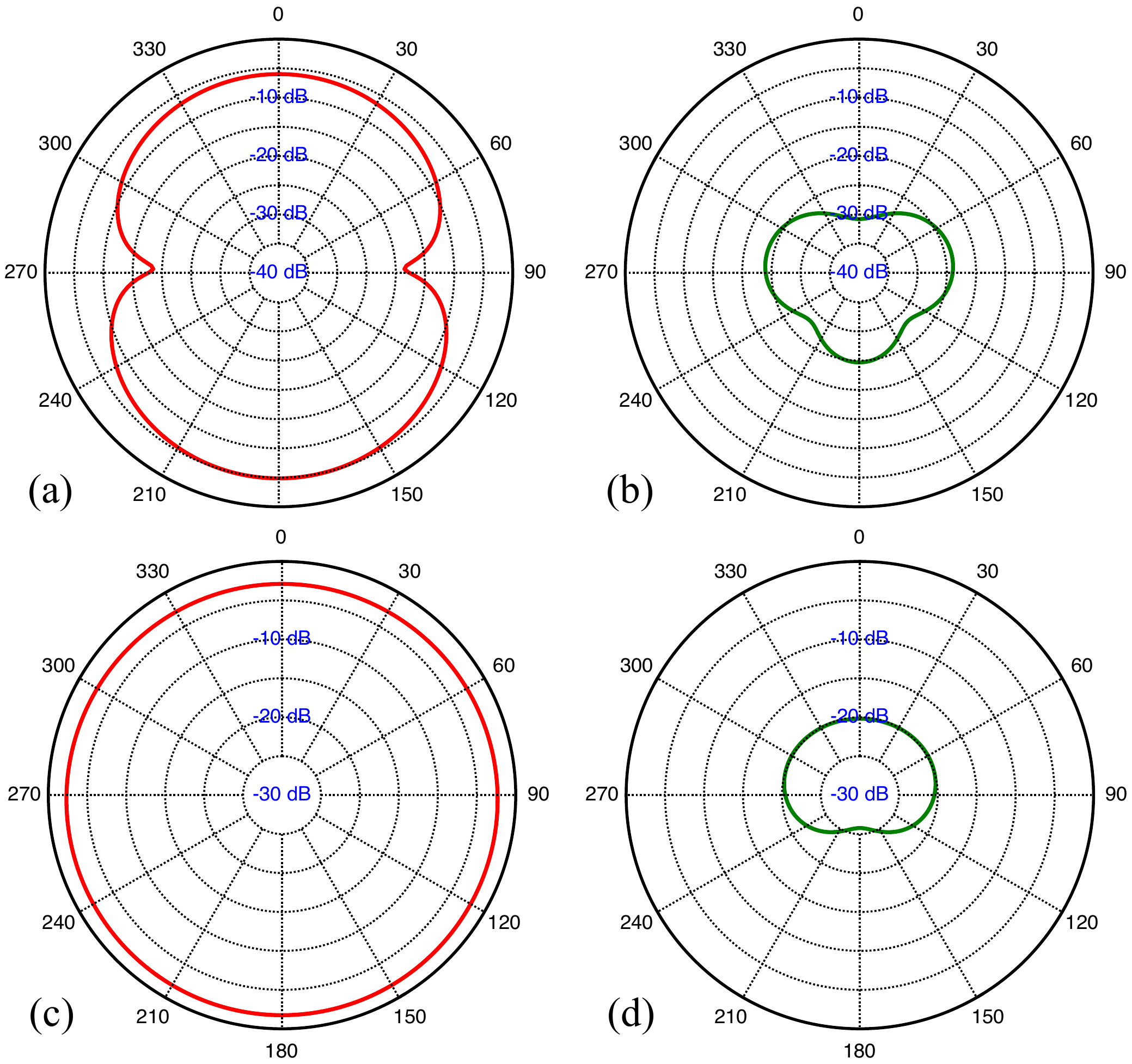}
\caption{Far-field scattering amplitudes in polar coordinates, and in logarithmic scale (a) for the bare scattering object with $\mu_{s,1}'=15\textrm{cm}^{-1}$ and $\mu_{a,1}=0.023\textrm{cm}^{-1}$, (b) for the cloaked scattering object with $\mu_{s,2}'=2.1\textrm{cm}^{-1}$ and $\mu_{a,2}=0.028\textrm{cm}^{-1}$, (c) for the bare absorbing object with $\mu_{s,1}'=6\textrm{cm}^{-1}$ and $\mu_{a,1}=0.15\textrm{cm}^{-1}$, (d) for the cloaked absorbing object with $\mu_{s,2}'=7\textrm{cm}^{-1}$ and $\mu_{a,2}=-0.25\textrm{cm}^{-1}$.}
\label{fig4}
\end{figure}

In particular, the minimum of $\sigma^\textrm{scat}$ is obtained for the coordinates (7, -0.25) that fits as well with the predictions of Eqs.~(\ref{l=0}), and~(\ref{l=1}). By comparing Figs.~\ref{fig3}(a) and~\ref{fig3}(b), we can notice also that in the first case, sensitivity to $\mu_{s,2}'$ is more evident, since the object is of scattering nature, whereas in the latter scenario, we can notice more sensitivity to $\mu_{a,2}$. This can be clearly seen from the shape of the spots (elongated in the $y$ and $x$ directions, respectively). It is interesting to obtain, from numerical simulations, taking into account many scattering orders, huge scattering reduction, sensibly around the same coordinates, with values of almost -45 dB. The quasistatic analysis, considered in our contribution, shows the importance of considering the effect of both the scattering and absorption coefficients of the shell. 

\subsection{Far-field SCS}

To further illustrate the capability of the proposed SCT cloaks for DPDW, far-field scattering patterns are shown in Figs.~\ref{fig4}(a)-\ref{fig4}(d). These plot the scattering amplitude in polar coordinates, in the $x-y$ plane [due to the symmetry of the structure as can be seen in Fig.~\ref{fig1}(a)]. The bare scattering and absorbing far-fields are given in Figs.~\ref{fig4}(a) and~\ref{fig4}(c), respectively, and significant scattering can be noticed. The cloaked structures, using the optimal parameters deduced from Fig.~\ref{fig3} are given for comparison in Figs.~\ref{fig4}(b) and~\ref{fig4}(d), respectively. These plots show that both types of heterogeneities can be made almost undetectable at all angles.

\begin{figure}[b]
\centering\includegraphics[width=0.85\columnwidth]{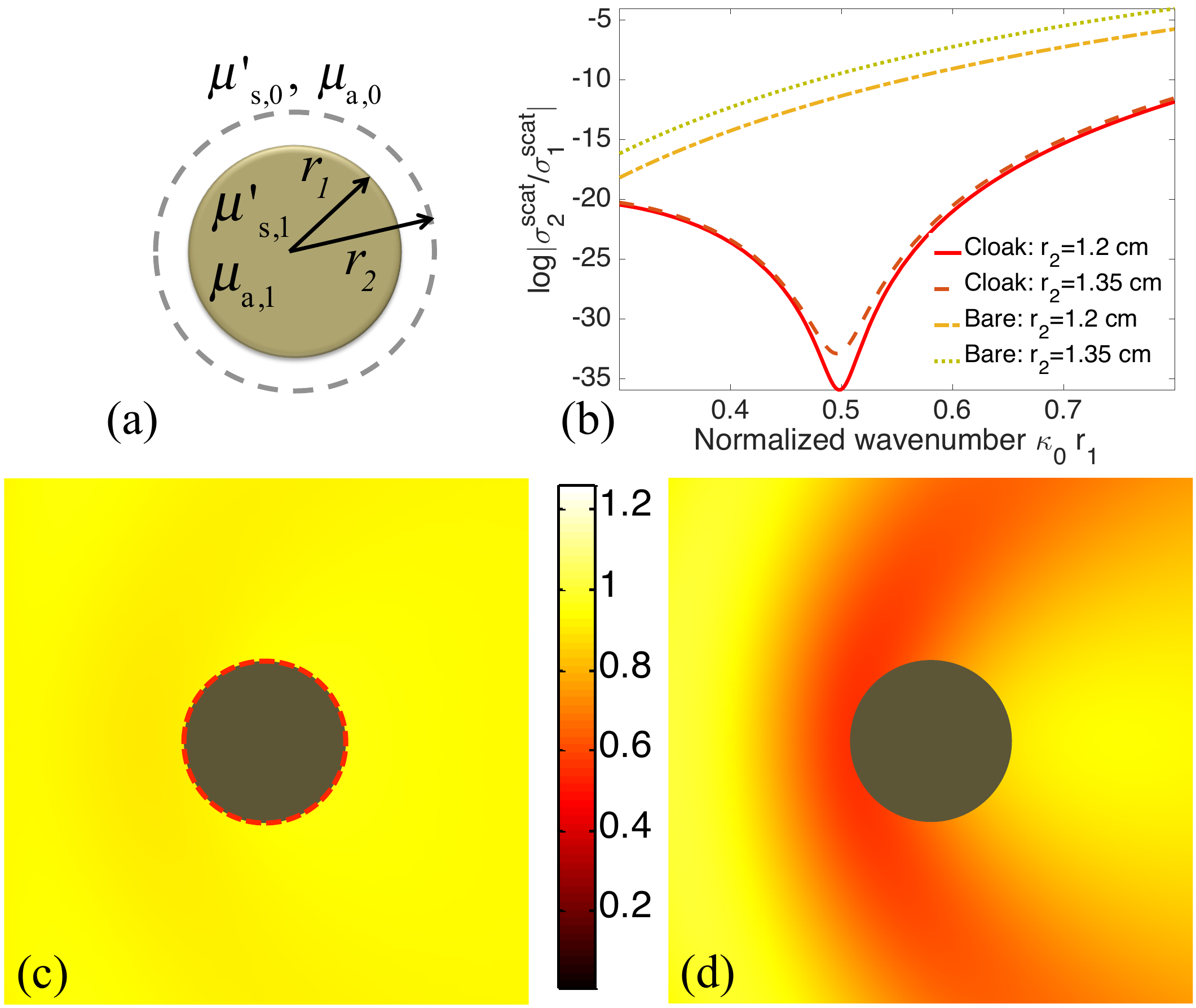}
\caption{(a) Scheme of the mantle cloaking structure with the dashed grey line showing scattering object ($\mu_{s,1}'=15 \textrm{cm}^{-1}$ and $\mu_{a,1}=0.023 \textrm{cm}^{-1}$) in the same turbid medium as in previous sections, and the ultra-thin shell with tailored impedance. (b) Normalized SCS of the structure given in Fig.~\ref{fig5}(a) versus normalized wavenumber $\kappa_0r_1$ for two radii of the mantle and bare objects with same radii, for comparison. Amplitude of DPDW for (c) the mantle-cloaked object and (d) bare scattering object.}
\label{fig5}
\end{figure}

\subsection{Mantle cloaking}

Figure~\ref{fig5}(b) reports the dependencies of the scattering cross section on the normalized wavenumber $\kappa_0r_1$ of the mantle cloak [schematized in Fig.~\ref{fig5}(a)] for two radii of the mantle (conformal: $r_2=1.2 \textrm{cm}$ and $r_2=1.35 \textrm{cm}$) where the bare spheres with similar radii (gray-dashed line) are also shown for a fair comparison. We suppose in this calculation that the surface reactance $X_d$ does not vary with frequency (this may be a good approximation around a range of wavenumbers around $\kappa_0r_1$). The uncloaked scenario (long-dashed line) and an uncloaked particle of radius $r=r_2$ is also drawn for a fair comparison. It is evident that an excellent scattering reduction can be achieved over a large range of frequencies for both cases. Figures~\ref{fig5}(c),~\ref{fig5}(d) map of the amplitude of the DPDW diffusing at a time instant scattered by a mantle-cloaked, uncloaked scattering obstacle [with same parameters as in Fig.~\ref{fig3}(a)], respectively. When it is surrounded by the  cloak, both forward and backward scattering are nearly vanished, with not too much difference between the DPDW diffusing in homogeneous medium and the wave bent by the cloak. This reduction of scattering is achieved thanks to the proper choice of the surface diffusive impedance (reactance), which restores almost uniform amplitude all around the cloak.

\section{Summary} 

To conclude, we have studied analytically and numerically cloaking mechanism based on the scattering cancellation technique in the context of diffuse photon density waves. In particular, we have shown that both a cloaking shell with convenient diffusivity and photon lifetime or mantle shell with tailored impedance, can drastically reduce the scattering from a spherical obstacle. We believe that our results help making the cloaking theory one step closer to its practical realization for diffusive light. This mechanism can also be used to build noninvasive medical imaging devices (e.g., ultrasound imaging) with moderately broadband features.

Finally, Our analysis of the light diffusion equation carried out in this paper has unveiled some new mathematical and physical features of DPDW cloaking, unseen in the context of heat diffusion, but also in acoustic, optical or mechanical waves. Geometric transform based designs of light diffusion cloaks should thus lead to a whole new range of exciting effects that would require specific theoretical, numerical and experimental investigation.

\end{document}